\documentstyle[prl,aps,epsfig]{revtex}
\begin{document}
\title{Scales of the Extra Dimensions and their Gravitational Wave Backgrounds}
\author{Craig J. Hogan}
\address{Astronomy and Physics Departments, 
University of Washington,
Seattle, Washington 98195-1580}
\maketitle
\begin{abstract}
Circumstances are described in which symmetry breaking during the formation of our 
three-dimensional brane within a higher-dimensional space
in the early universe   excites mesoscopic classical radion
or brane-displacement degrees of freedom and produces a  detectable stochastic background of
gravitational radiation.  The   spectrum of the  background 
is related to  the unification energy scale and the the sizes and numbers of large extra
dimensions.  It is shown that properties of the  background  
observable  by gravitational-wave observatories  at frequencies 
$f\approx 10^{-4}$ Hz to $10^3$ Hz contain  information about
unification on energy scales from $1$ to  $10^{10}$ TeV,
gravity propagating through  extra-dimension sizes  from 1 mm to  $10^{-18}$mm,
and   the dynamical history and stabilization of from one to seven extra dimensions.
\end{abstract}
\section{Relic Backgrounds from Brane Formation in ``the   Desert''}

If  quantum gravity  lies fundamentally
 in ten   spatial dimensions,  then   the
 seven  ``extra'' dimensions must somehow be hidden.
The traditional  approach, going back to Kaluza and Klein,
is to make extra dimensions very small, close the Planck scale.
 Recently there has been a resurgence
of interest in another approach where many of the extra dimensions are of 
 much larger
size\cite{arkanihierarchy,antoniadas,arkaniphenom,randallsundrum1,randallsundrum2,lykkenrandall}. 
The Standard Model fields are confined to a three-dimensional wall or ``brane'', a 3-D defect in the
higher-dimensional space; only gravity can propagate in the other dimensions or ``bulk''.

These ideas are observationally constrained   from the 
particle-physics side and from the gravity side.    Roughly speaking, the 
success of the Standard Model
limits the   unification energy scale to  energies well above  1 TeV,
although some of the most interesting brane+bulk scenarios are those 
where the unification scale is not many orders of magnitude more than a TeV, providing a
solution to to the ``hierarchy problem.''
 Laboratory
measurements of gravity limit the size (or radius
of curvature) of the extra dimensions to be smaller than about 0.3 mm\cite{adelberger}.   
Constraints of a model-dependent nature, including astrophysical ones,  
extend these limits, since the models generally add
new particles and interactions to the Standard
Model (e.g.,\cite{arkaniphenom,goldbergerwise,kaplanwise}). A    broad range of unification scales
and brane configurations  is consistent with current constraints; the unification scale and extra
dimension size can be anywhere between the TeV/mm limits  and the Planck scale. 

This broad range of possibilities 
of course just reflects the relative lack of data in the physics/cosmology
``desert''. Standard unification ideas often show running couplings extrapolated
from actual experiments at 100 GeV
by 14 orders of magnitude to the unification scale at $10^{16}$ GeV, but there
are few direct particle data beyond the TeV scale.  
A similar lacuna
appears in the  cosmological data; we  can
cite direct information from  cosmic abundances  about microphysical processes
as early as 1 MeV, which can be affected by inhomogeneities originating
as early as 100 GeV (e.g.
\cite{kainulainen}) and about earlier inflationary effects on very large scales (such the microwave
background anisotropy, e.g. \cite{lange}), but we have   almost no direct data
about early ``mesoscopic'' structure on comoving scales smaller than a few centimeters
at 100 GeV (a quantity of energy about equal to the mass of the Earth)  at times
earlier than $10^{-9}$ second (except possibly for the mean  baryon-to-photon ratio
 and  the mean density of  some forms of  dark matter.)
 Scalar metric
 perturbations are very effectively erased on these small scales by neutrino
diffusion long  before they can have any effect on observables such
as element abundances\cite{heckler}. Small-scale perturbations in baryon/photon ratio
$\eta$ on small scales, even  of large amplitude,
 are erased by nucleon
diffusion before nucleosynthesis\cite{applegate,malaney,kainulainen}.  From   most of 
cosmic history (in
log space) since the Planck time, and on most scales of structure, only the
tensor modes--- gravitational waves---
survive to the present.

New gravitational wave observatories
now under development will soon achieve a critical level of sensitivity
at frequencies
where they can plausibly detect the relic   stochastic
background of gravitational waves from  cosmological events far into 
the    early, previously unobservable ``mesoscopic'' era. One of the most interesting possible
sources of radiation, from the point of view of both physics and cosmology,
is the formation of our 3-brane.
This paper surveys broadly the circumstances and the
  values of parameters--- the number  and the 
size of large extra dimensions, and the scale of unification---   for
which the background  might be detected with current technology.

To create a detectable background, the metric during the unification era
must be far from equilibrium--- it must be close (within a few orders of 
magnitude) to chaotic on the mesoscopic
scale of the horizon at unification. Such circumstances can indeed be more generic
than the usual assumption of a uniform equlibrium system. 
The  brane+bulk  configurations being contemplated are classical setups: 
  the vacua of the fundamental theory have effective potentials
that tend to drive them to   forming classical three-dimensional defects.
The details of the stabilization of the setup are not known
  but the cosmological formation of such defects is 
in general not an equilibrium achieved by a microscopic quantum process--- it involves macroscopic
(or at least mesoscopic) collective dynamics, in the same way that 
spontaneous symmetry breaking occurs in the standard Higgs mechanism. In the cosmological
context, symmetry breaking during
 the setting up of the brane is likely  to create  gravitational waves via mechanisms
analogous to such intense classical sources as cosmic strings or first-order
phase transitions. The properties of the waves can be informative about  both the 
physics of the brane and  its early cosmological history.

\section{Characteristic   Frequency and Amplitude}

 The formation and 
stabilization of the brane are associated with two new
geometrical degrees of freedom,  radion modes governing the size
or curvature of the extra dimensions and Nambu-Goldstone modes corresponding
to spatial variations in
brane displacement\cite{hoganprl}.
The characteristic  scales  of the resulting broadband background spectra can   be estimated
from general scaling considerations.
 The characteristic  amplitude  of the metric perturbations induced by
brane condensation is of the the order of unity and the 
characteristic frequency is   the horizon scale. We will use the 
``maximal  amplitude'' and ``Hubble frequency'' to estimate the extra-dimensional scales accessible 
to gravitational wave detectors.

  The characteristic scale for setting up our 3+1 world  is  the gravitational timescale 
$H^{-1}$  in
standard 3+1  General Relativity.
This is cleanly determined by General Relativity and thermodynamics
except for    a weak dependence on  the
particle-physics uncertainties encapsulated in   the  number of effective
relativistic degrees of freedom $g_*$  (e.g.\cite{kolb}). 
The characteristic frequency of observed 
gravitational waves is thus\cite{hogangwb,maggiore,maggiore2}  the Hubbble frequency 
  redshifted to
the present day,   $f_{H0}(T)\equiv H(T) a(T)/a_0$: 
\begin{equation}
f_{H0}=7.65\times 10^{-5}\ {\rm
Hz}\ (T/TeV)g_*^{1/6}(g_*/g_{*S})^{1/3}T_{2.728}=
 9.37\times 10^{-5}\ {\rm Hz}(H\times 1{\rm mm})^{1/2}g_*^{-1/12}(g_*/g_{*S})^{1/3}
T_{2.728}.
\end{equation}
The 
estimate  is valid back to the  threshold of the extradimensional dynamics:
$H^{-1}\le b_0$, (where $b_0$ is the 
size or curvature radius of the largest extra dimension), or $T\ge M_*$, whichever comes first.

The fiducial ``maximal'' energy density is set\cite{hogangwb,maggiore,maggiore2} by 
the mean energy density in  all relativistic species (photons and three massless neutrinos),
$\Omega_{rel}=8.51\times 10^{-5} h_{70}^{-2} T_{2.728}^4$,
where $h_{70}$ refers to the Hubble constant.
 The energy density of relativistic matter redshifts in the same way as  gravitational waves
so they     have the same ratio of  energy densities today as when the 
waves were generated. (Constraints
from nucleosynthesis limit the gravitational waves to about 10\% or less 
of the relativistic matter density, or rms strain about three times smaller
than the maximal value.)

At the low frequencies  observed by spacecraft interferometers (such as LISA) there are many
astrophysical foregrounds, including known sources such as galactic binaries.  
A stochastic background 
can however  be distinguished from other sources of noise and astrophysical wave sources
by resolving the background in frequency.
For the projected sensitivity of LISA, a  maximal background is detectable
 above the instrument noise
or the other likely  astrophysical foregrounds\cite{hoganprl,lisa98,armstrong} over a frequency
range from about $10^{-1}$ to $10^{-4}$Hz. In the center of this range,
 $\approx 10^{-4}$Hz, backgrounds are detectable with
\begin{equation}
\left({\Omega_{GW}(\Delta f = f)\over\Omega_{rel}}\right)\approx 10^{-6}.
\end{equation}

At the higher frequencies observed by ground-based
observatories  (e.g., LIGO, VIRGO, TAMA300, GEO600) it could well be that
 other stochastic astrophysical backgrounds are relatively weak, and that the 
observable sources are so limited in time duration and frequency that   the main
contribution to the rms noise comes from  the instrument itself.
 In this case
we can use the entire bandwidth  to measure the background and reach a level which
places meaningful constraints on  backgrounds\cite{maggiore,allen}. 
The most promising technique is to correlate the signals from
two interferometers within a wavelength  of each other. 
  Combining two early LIGO
or VIRGO systems will likely achieve a sensitivity\cite{maggiore} 
\begin{equation}
\left({\Omega_{GW}(\Delta f = f)\over\Omega_{rel}}\right)\approx 10^{-2},
\end{equation}
and LIGO II may reach another two orders of magnitude below this\cite{kip}---
 close to the level where other stochastic astrophysical
 backgrounds (such as neutron star emission) may be expected.  
Sufficient sensitivity to constrain primordial stochastic backgrounds
 may be expected for early ground-based interferometers roughly from 50 to 500
Hz.

\section{Excitation of Classical Displacement and Radion Perturbations}

There are many possible ``brane worlds'' with various configurations of
large extra dimensions.
 To simplify the discussion we will assume that there are
$n$ equal-size largest extra dimensions of size $b_0$ and all the others
are much smaller, of size $M_*^{-1}$ where $M_*$ is the true fundamental
unification scale.   
 The apparent
(usual) Planck mass in 3+1-D, $M_{Planck}$, is approximately given by 
$M_{Planck}^2\approx M_*^2 (M_*^nV_n)$ where 
$V_n$ is the volume of the extra dimensions. 
In our situation, 
\begin{equation} 
M_{Planck}^2\approx M_*^2 (M_*^nb_0^n)
\end{equation}
where $b_0^n$ is the volume of $n$ extra large dimensions.
Thus each choice of $n$ defines a relation between $M_*$ and $b_0$,
shown in Figure \ref{fig: variousn}. The various relations intersect at $M_*=M_{Planck}$:
\begin{equation} \label{eqn: gauss}
b_0\approx (M_{Planck}/M_*)^{2/n}M_*^{-1}.
\end{equation}  
Gravity is ``normal''\footnote{Gauss' law shows that gravity obeys a $r^{-2}$ law for $r>b_0$, and
$r^{-2-n}$ for $r<b_0$.} on scales larger than $b_0$; standard field
theory holds on the brane at energies below $M_*$.

The cosmological  formation of the brane setup is regulated on 
the gravitational timescale. 
In the 3+1-D era,   this is just given by the   Hubble scale  discussed above,
\begin{equation}  
H^{-1}/ 1{mm}\approx (T/ 1{TeV})^{-2}
\end{equation}
 which  defines the usual gravitational  relationship between 
length and energy. If we follow 3+1-D cosmology back using this equation,
we come first  to either (1) the time when $T=M_*$  and extra-dimensional effects
come into play on a small scale, possibly associated with a first-order phase 
transition; or (2) the time when $H=b_0^{-1}$, and the geometrical
degrees of freedom  of the extra dimensions are important even 
on large scales.
In general the excitation of the extra-dimensional modes has a different
character depending on whether $b_0(M_*)$ lies above or below
the Hubble relation $H(T=M_*)$.
The former case produces displacement modes at with a scale determined by $b_0$;
the latter case produces radion modes with a scale determined by $M_*$.

 The Hubble  relation $H(T)$ is degenerate with the 
$n=2$ relation between $b_0$ and $M_*$, meaning that in 
this case  macroscopic and microscopic
departures from the standard picture happen at about the same time. The
transition to  new
microscopic unification physics   and to  5-dimensional cosmological
expansion would both happen at the same point on these lines, also corresponding to the
condensation of the brane and the creation of the gravitational waves.

 In the $n=1$ case where $b_0>H(T=M_*)^{-1}$, the 
condensation of the brane (at  $T=M_*$) takes place while the
Hubble length is still much smaller than $b_0$. In this case 
a long evolution takes place after  the brane forms
(and non-gravitational fields are confined to 3D as today),
during which  the cosmological evolution involves one   classically large
extra dimension and the usual relativistic cosmology does not apply. 
In particular there is not time for signals to propagate as far as $b_0$
in an expansion time. If the formation of the brane spontaneously breaks
Poincar\'e invariance of the whole spacetime, 
 the  position of the  brane when it condenses is  therefore uncorrelated 
on scales larger than $H(T=M_*)^{-1}$ and  less than $b_0$. The
Nambu-Goldstone modes corresponding to brane displacement are therefore 
substantially excited by the Kibble mechanism\cite{zeldovich,kibble} up to
wavelength $b_0$.  These large-amplitude scalar perturbations
dynamically couple to the tensor modes to produce a gravitational wave background,
in much the same way that gravitational waves are generated by defects confined
to a 3+1-D space\cite{hoganprl,hogangwb,allen,gold82,krauss,gold95}.
The spectrum is peaked 
at the corresponding redshifted Hubble frequency $f_{H0}(H^{-1}=b_0)$
and falls off as a power law at higher and lower frequencies. The  spectrum
at high frequencies depends on details of the $4+1$-D era of cosmic evolution, but
may be as strong as $\Omega_{GW}(\Delta f=f)= $ constant; at lower frequencies, it is 
estimated\cite{hoganprl} to fall off as $\Omega_{GW}(\delta f=f)\propto f^{7}$.

  In the $3\le n\le 7$
case where
$b_0<H(T=M_*)^{-1}$, the universe can find  itself in a macroscopic 
space of
conventional 3-D dimensionality (on the Hubble scale)   at  the
temperature   where new extra-dimensional unification physics  comes into play on
a microscopic scale. Conventional cosmology can therefore be used
as a framework.  The stabilization of the $3\le n\le 7$ extra dimensions 
appears as a change of physics, nongravitational and gravitational,
which can be described macroscopically  as a  change in the energy-momentum tensor
of cosmic matter, and  may be (3D-) spatially  inhomogeneous as a result of 
symmetry breaking. The  excitation of
these ``radion modes'' can be described as a   change in the order parameter of a
vacuum state  in 3D, as is familiar in cosmological phase
transitions. If this transition described by the effective radion potential
is first order, free energy is 
released in macroscopic flows created by bubble nucleation and collisions,
and   can have a substantial
gravitational-wave
 component\cite{transition84,transition86,transition94,gleiser}, although
for most plausible situations these backgrounds are several orders  of
magnitude weaker than $\Omega_{rel}$. Figure \ref{fig: variousn}  
illustrates  that the phase-transition bubbles are smaller than
the horizon and the peak of the spectrum  
lies significantly  higher than the redshifted Hubble frequency.\footnote{Account is
not taken of the narrower instrumental bandpasses for the less-than-maximal 
intensity of the background in these cases.}

\section{Range of accessible parameters}

\begin{figure}[htbp] 
\centerline{\epsfig{height=5in, file=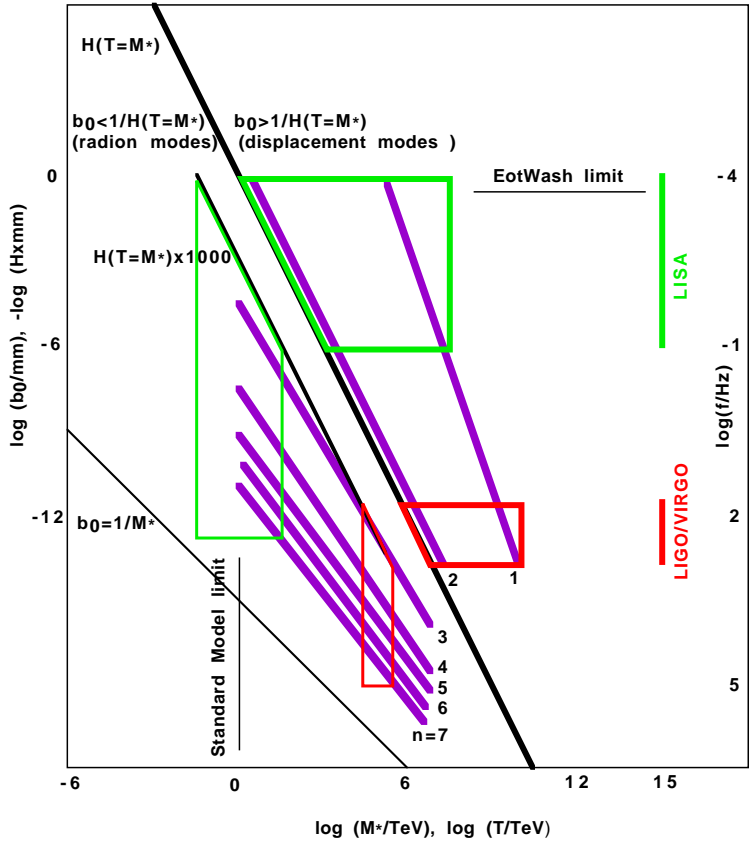}}
\vspace{0.25in}
\caption{ \label{fig: variousn}Summary of the new parameter space of extra dimensions that will be
probed by gravitational-wave interferometers. In the context of models where $n$ extra dimensions
have the same size
$b_0$ and the rest lie at the unification scale $M_*^{-1}$,
a choice  of $n$ defines the  relation  between   $M_*$ and   $b_0$; these relations
are shown  labeled by $n$. Classical, 3+1-D relativistic cosmology
(the Friedmann equation) defines the 
 relation $H(T=M_*)$ as shown. Theories with $b_0\ge 1/H(T=M_*)$   
condense their classical 3-brane before the Hubble  length is as large as $b_0$,
and the random location of the brane
in the extra dimensions   excites Nambu-Goldstone displacement modes; the rms
strain peaks near the frequency $f_{H0}(H=b_0^{-1})$ and is damped at lower frequencies.
Theories with $b_0\le 1/H(T=M_*)$ on the other hand are already 3+1-D on the Hubble scale
when they pass the unification temperature; the final stabilization of 
the extra dimensions then appears as a phase transition in a 3+1-D cosmology,
with an order parameter represented by the radion. The strain in this case peaks
at a frequency determined by the nucleation scale, typically at least 100 times
higher than $f_{H0}(T=M_*)$; the relation $H(T=M_*)\times 1000$
(or $f=1000\times f_{H0}(T=M_*)$) shows the peak
frequency for an illustrative radion background.
 Boxes indicate the corresponding regions of these
parameters which may give rise to   detectable mesoscopic gravitational radiation backgrounds
in the LISA and LIGO bands.
Heavy-line boxes show the displacement mode parameters, lighter-line
boxes show the radion mode parameters.  These regions extend well beyond those already
constrained by gravitational experiments, direct particle production, or other astrophysical 
constraints. Theories which ``solve the hierarchy problem'' have $M_*$
close to the Standard Model limit, and all of the viable ones ($2\le n\le 7$)
could possibly produce an observable background of one type or the
other in the LISA band.}
\end{figure}

Although   based on a simplified picture,
 Figure \ref{fig: variousn} illustrates accurately the 
 wide range of $n,b, M_*$ potentially accessible 
via this technique. 
The possibilities include both the   radion and Nambu-Goldstone displacement modes,
and include, for some  $M_*$ and $b_0$, all likely values of $n$. The observatories 
 constrain theories
with values of these parameters including  the limits accessible by 
other means, near  $M_*=1$  TeV  and  $b_0=1$  mm,  and extending beyond this range by
many orders of magnitude.  For example, direct laboratory 
gravitational experiments become extremely challenging below the mm scale, whereas
primordial gravitational wave backgrounds at the highest
detectable frequencies (about 1 kHz)   probe new classical spacetime dimensions  directly
(via the displacement modes)
on scales   14 orders of magnitude
smaller--- aided by the cosmological redshift which stretches the waves to detectable wavelengths.
Even a nondetection of backgrounds will constrain the behavior of theories with 
these numbers and sizes of extra dimensions.

Of particular interest are the models
 which potentially solve the hierarchy problem, where $M_*$ is not many orders of magnitude
more than 1 TeV.  For   these
models, backgrounds in the LISA band can arise from    all the viable numbers of extra dimensions
($n=2$ to
$n=7$), and a large range of dimension sizes---
$10^{-12}{\rm \ mm}\le b_0\le 1$mm,  depending on $n$.   (It is already known that the
$n=1$ case requires     a unification scale
of at least $10^6$ TeV to be consistent with the EotWash limits).  

The LIGO/VIRGO frequencies  
probe much higher energy unification scales, from  
$M_*\approx 10^4$ to $10^{10}$ TeV, and $b_0$ less than $10^{-10}$mm. 
The  intermediate 
range  of $b_0$ and  $M_*$, corresponding to 
the frequency gap between these two sets of experiments,   would be 
obervable with a   space interferometer with a smaller baseline 
than LISA. 

The smallest dimension accessible is determined by the highest frequency, which is 
set by the photon noise limit of LIGO or its successors. It is reasonable to expect
high sensitivity with this technology up to about  1 kHz, which probes
displacement modes down to 
$b_0\approx H^{-1}\approx  10^{-14}$mm. The largest $M_*$ probed (corresponding to $n=1$)  
can be as high as   about $10^{10}$ TeV; the smallest $b_0$ (corresponding to $n=7$)
can be as short as $10^{-18}$ mm.
There are no known strong 
astrophysical sources of gravitational radiation
above $10^4$Hz,   so 
new  types of experiments   at
higher frequency,  
 with   sufficient   sensitivity  (i.e. $\Omega_{GW}\le
\Omega_{rel}$),   would   probe   even more extreme 
physics with little astrophysical contamination. 

As an aside, it is interesting to consider the traditional case   with all the extra
dimensions  at the classical Planck scale. This point corresponds to $M_{Planck}^{-1}\approx
b_0\approx M_*^{-1}$ for 
any value of $n$. 
(The relation $b_0\approx M_*^{-1}$ shown in 
Figure \ref{fig: variousn} for other values of $M_*$ corresponds
to the $n=\infty$ limit of equation \ref{eqn: gauss}.) The gravitational wave background
in this model has a peak frequency of about  $10^{12}$ Hz, or a peak wavelength of about
0.3 mm! Although the maximal classical background typically  has
$10^{10}$  times more energy density than tensor perturbations from inflationary
 quantum effects,\footnote{Most discussions of primordial tensor modes 
concentrate on waves excited by quantum effects at inflation.
These much weaker  waves are probably unobservable at LIGO and LISA frequencies, 
but may already be   observable on much larger scales through their effect on microwave background
anisotropy. 
A gravitational wave background is produced by quantum fluctuations
of the graviton during inflation, with rms tensor-mode amplitude (e.g. \cite{kolb}) $h_t\approx
(H_{inflation}/M_{Planck})^2$. This is  most easily detected on
 comoving scales close to the present-day Hubble length, $f\approx H_0\approx 2\times 10^{-18}$ Hz,
where background radiation anisotropy implies an amplitude $h_t\le 10^{-5}$. Of course the amplitude
might be very different on the scales of the $10^{-4}$ to 1000 Hz waves considered
here, but since $H_{inflation}$ generally decreases slowly during inflation
the ``tilt'' from a scale-free spectrum generally makes
 $h_t$  smaller at higher frequencies. The
energy density of this background is $h_t^2$ times smaller than the maximal value ($h_t=1$) and
is likely to be unobservable.}
  there is
no plausible technique for detecting a gravitational wave background
at such high frequencies.

In models enclosed by the boxes in Figure \ref{fig: variousn},
properties of  these backgrounds can in some circumstances provide measurements of the main 
parameters of the largest extra dimensions: 
$n$, $b_0$ and $M_*$. 
For example, suppose that a stochastic background is detected with various signatures (such as
 anisotropy statistics)   leading us to suspect that it comes from high redshift. 
If the background is within one or two orders of magnitude of $\Omega_{rel}$ it is
likely from a displacement mode. This in turn indicates that $n=1$ or $n=2$--- one
or two extra dimensions are much larger than the rest. 
Since only $n=1$ undergoes a substantial period of classical
expansion with $b_0>H^{-1}$, the background spectra in these two situations
are likely to be qualitatively different:  the $n=2$ spectrum is more sharply 
peaked at $f_{H0}$ and the  $n=1$ spectrum includes a more intense high-frequency
tail from modes excited before $H_0^{-1}=b_0$. Knowing $n$, the frequency of  the spectral peak then
provides an estimate of
$b_0$ and
$M_*$.  A nondection of a displacement-mode background conveys a significant constraint 
on parameters and on cosmology since the conditions for generating a background
from    
symmetry breaking in this regime are relatively  generic. 

In the case of the $n\ge 3$ radion modes,  it is difficult to break
the degeneracies using just the information available from  the background measurements.
For example, the peak frequency and intensity depend on the details of bubble
nucleation dynamics as well as on $f_{H0}$; these in turn depend on the details
of the effective radion potential.  That is, a weaker transition 
leads to a  smaller nucleation scale, a  higher frequency peak and a less intense
background. 
 (The interpretation of the background might  
in the  case of LISA-band backgrounds
 be supplemented by other data, since the energy scale $M_*$ is 
not too far above energies accessible by direct experiment.) A nondetection in
this regime will only constrain the corresponding parameters   in particular models;
lack of a background could just mean that the relevant
transition is not strongly first order, but occurs gently and does not
 generate intense gravitational waves. 

\acknowledgements
I am grateful for useful discussions with E. Adelberger, B. Allen,
 B. Barish, P. Bender, D. Kaplan,
A. Nelson, and
 K. Thorne.
I thank  
the  Albert Einstein Institute, Golm, the LIGO Hanford Observatory, and JPL for hospitality.
This work was supported at the University of Washington
by   NSF.
{}

\end{document}